\documentclass[nofootinbib, twocolumn,showpacs, showkeys,superscriptaddress,10pt]{revtex4-2}
\usepackage{amsmath,amssymb,graphicx, subfigure, hyperref, braket, float, relsize}
\usepackage{blindtext}

\counterwithout{figure}{section}
\usepackage[export]{adjustbox}
\graphicspath{{figure/}}
\usepackage[T1]{fontenc}
\usepackage{ae,aecompl}
\DeclareUnicodeCharacter{0301}{\'{e}}

\usepackage{graphicx, xcolor}
\begin{document}

\title{Effect of induced transition on the quantum entanglement and coherence in two-coupled double quantum dots system}

\author{Zakaria Dahbi} \email{zakaria\_dahbi2@um5.ac.ma}
\address{Lab of High Energy Physics-Modeling and Simulation, Faculty of Sciences, \\Mohammed V University in Rabat, 4 Avenue Ibn Battouta B.P. 1014 RP, Rabat, Morocco}
\author{Maron F. Anka} \email{maronanka@id.uff.br}
\address{Instituto de Física, Universidade Federal Fluminense, Av. Gal. Milton Tavares de Souza s/n, 24210-346 Niterói, Rio de Janeiro, Brazil}
\author{Mostafa Mansour}\email{mostafa.mansour.fpb@gmail.com}
\address{Laboratory of High Energy Physics and Condensed Matter, Department of Physics,\\ Faculty of Sciences of Ain Chock, Hassan II University, P.O. Box 5366 Maarif, Casablanca 20100,  Morocco}

\author{Moisés Rojas} \email{moises.leyva@ufla.br}
\address{Departamento de Física, Universidade Federal de Lavras, 37200-900, Lavras-MG, Brazil}

\author{Clebson Cruz} \email{clebson.cruz@ufob.edu.br}
\address{Grupo de Informação Quântica e Física Estatística, Centro de Ciências Exatas e das Tecnologias, Universidade Federal do Oeste da Bahia—Campus Reitor Edgard Santos. Rua Bertioga, 892, Morada Nobre I, 47810-059 Barreiras, Bahia, Brazil}
\begin{abstract}
Studying quantum properties in solid-state systems is a significant avenue for research. In this scenario, double quantum dots (DQDs) appear as a versatile platform for technological breakthroughs in quantum computation and nanotechnology. This work inspects the thermal entanglement and quantum coherence in two-coupled DODs,  where the system is exposed to an external stimulus that induces an electronic transition within each subsystem. The results show that the introduction of external stimulus induces a quantum level crossing that relies upon the Coulomb potential changing the degree of quantum entanglement and coherence of the system. Thus, the quantum properties of the system can be tuned by changing the transition frequency, leading to the enhancement of its quantum properties.
\end{abstract}

\vspace{2cm}
\maketitle

\section{Introduction  \label{sec1}}
Studying theoretical quantum information quantifiers in condensed matter physics systems has paramount importance as a key resource for several information processing protocols \cite{R4,R5,MK,MK2,RQB,R1,R2,R7,R8,cruz}. In this scenario, double quantum dots (DQD) appear as a protean platform for condensed matter physics due to its role in the development of emerging quantum technologies \cite{donsa2014double,de2021two,elghaayda2022local, ferreira2022thermal,majek2021majorana,zhang2021anisotropic,ginzel2020spin,R43,R44,R45,R46}. Entanglement and quantum dynamics of two electrons within coupled DQDs were examined, and aspects of quantum coherence were addressed in Refs. \cite{R50, R51}. In general, quantum entanglement has attracted the attention of the scientific community in the past few years due to its pivotal role in developing new quantum-based technologies \cite{R4,R5,MK2,RQB,R1,R2,R7,R8}. 

Despite its enormous promise, measuring and characterizing entanglement in multipartite quantum systems is a rather complicated task \cite{hiesmayr2021free,roik2022entanglement}. Numerous research for characterizing the degree of entanglement in multipartite quantum systems have been conducted \cite{cruz2017,castro2021,R9,R10,R11,R12,R13,R14,R15,R16, dahbi}. However, entanglement does not encompass all quantumness in a system \cite{zurek,vedral}.  On the other hand, quantum coherence is derived from the quantum state superposition principle  \cite{xi2015quantum}. Similar to quantum correlations, it is a fundamental idea in quantum theory. It is also a crucial physical resource for quantum computations \cite{R32,R33,R34}. Quantum coherence is defined as the generalized quantum uncertainty in reference \onlinecite{R39} and is regarded as a useful and necessary resource in quantum information tasks \cite{R32,R40,R41, MK}. In order to investigate the relationship between quantum coherence and quantum correlations, a new measure has been introduced known as correlated coherence \cite{kraft2018,tan2016}. Correlated coherence measures the total amount of coherence stored in quantum correlations. The link between quantum coherence and correlations is discussed in references \onlinecite{R32, tan2018, R42}. In the scenario of DQDs, these connections between quantum coherence and other quantum resources have received considerable attention \cite{R52,R53,R51,R54,R55}. 

In this regard, this work studies the effect of an external stimulus on the quantum aspects of thermal coherence and entanglement in a two-coupled DQDs system. The external stimulus is characterized by the transition frequency of the electrons between the left and right dots. The physical model is introduced in order to provide the literature with a comparison between the total, local, and correlated $l_1$-norm quantum coherence with the amount of entanglement in the system. The influence of the transition frequency on these quantum resources is studied, typifying the degree of local and correlated quantum features in coupled DQDs systems and exploring the influence of the temperature, Coulomb coupling, and tunneling strengths. The results show that the competition between the Coulomb Coupling and the energetic contribution of the external stimulus leads to a quantum level crossing between a separable and an entangled configuration of the system, which leads to the enhancement of its quantum properties.

\section{Two coupled double quantum dots model} \label{sec3}
The proposed physical model comprises two sets of double quantum dots (DQDs), where each quantum dot can only contain a single electron. Each electron has two degrees of freedom, so it can either be in the left dot ($\ket{l_i}$) or in the right dot ($\ket{r_i}$). A sketched setup of the model is shown in Fig. \ref{DQDS}. An external stimulus is conveyed by an electromagnetic wave ($\gamma$) emitted from an external source. In reaction to the external stimulus, the DQDs system display variations in its quantum aspects concerning the energy absorbed by each electron, resulting in electron transitions between left and right dots.
\begin{widetext}
\begin{center}
\begin{figure}[ht]
\centering
\includegraphics[scale=0.53]{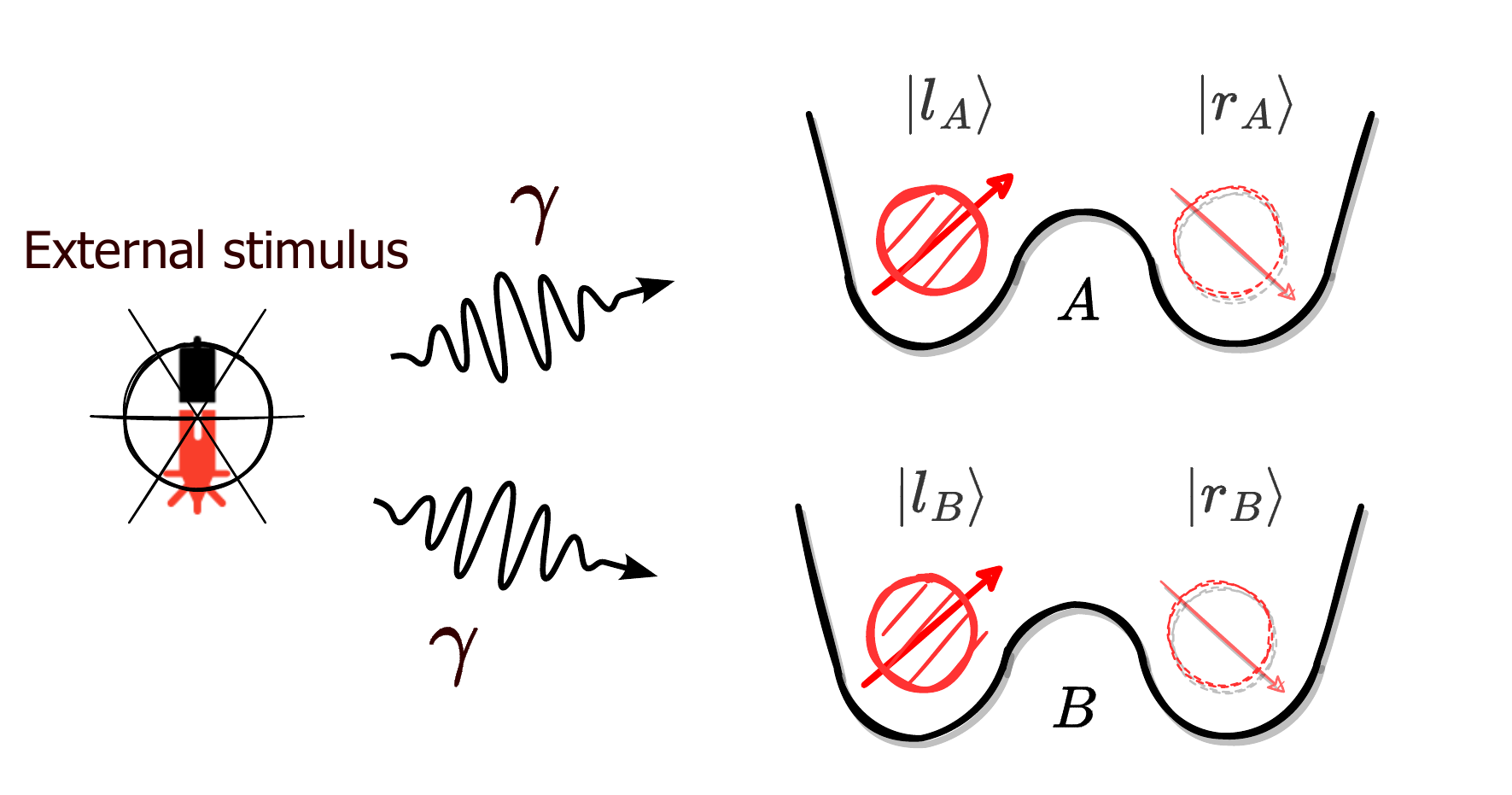} \includegraphics[scale=0.16]{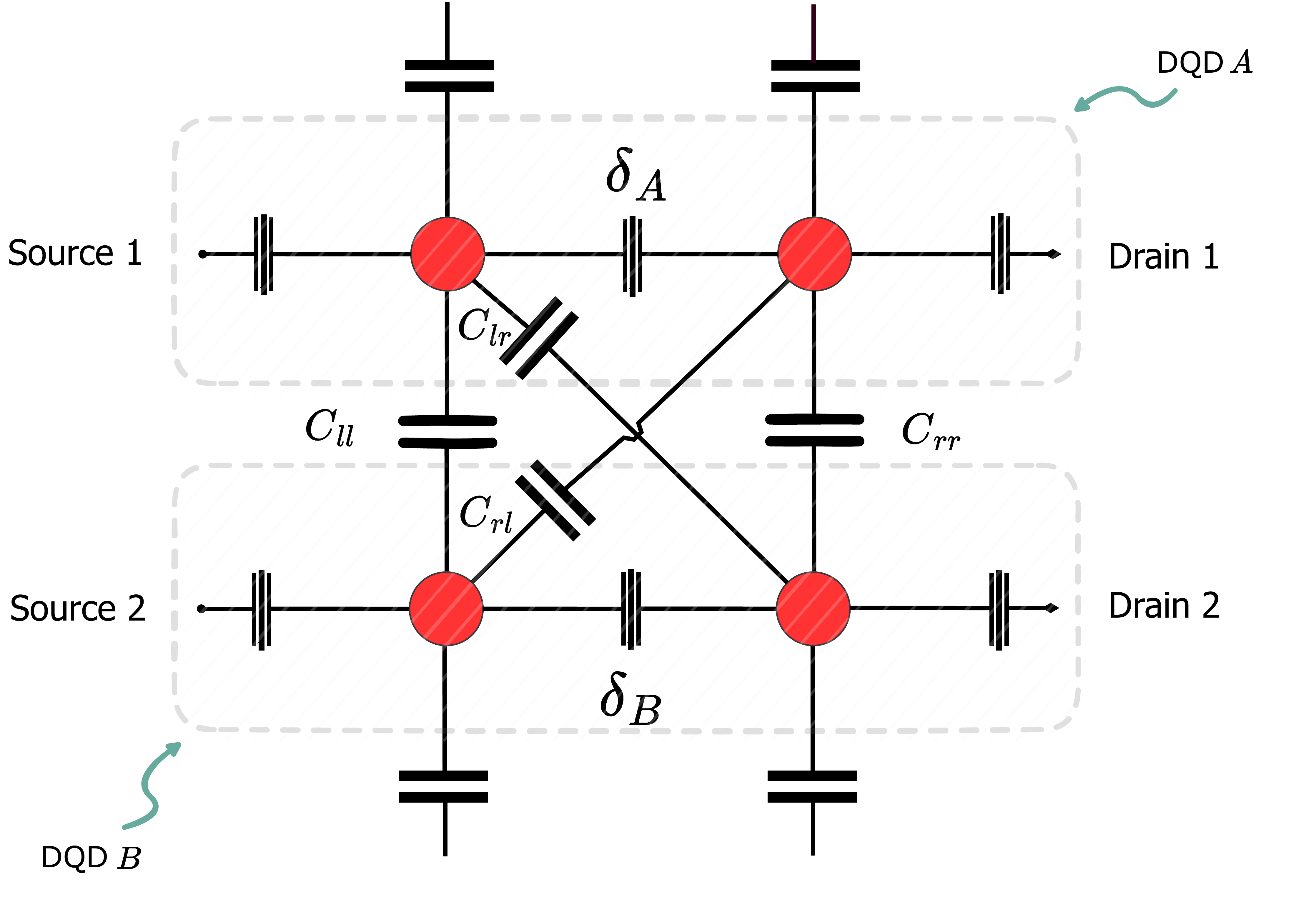}
\caption{A scheme of the two coupled DQDs model with an external stimulus (left) and its equivalent circuit (right). $C_{lr}$, $C_{rl}$, $C_{ll}$ and $C_{rr}$ are the capacitors connecting the DQDs.}
\label{DQDS}
\end{figure}
\end{center}
\end{widetext}
The Hamiltonian of the two coupled double quantum dots can be written on its explicit form in the local basis $\{\ket{i_{A},j_{B}}\}_{i,j=l,r}$ as
\begin{equation}
H=\left(
\begin{array}{cccc}
 \mathcal{V}+2 \hbar \omega & \delta_B & \delta_A & 0 \\
 \delta_B & -\mathcal{V} & 0 & \delta_A \\
 \delta_A & 0 & -\mathcal{V} & \delta_B \\
 0 & \delta_A & \delta_B & \mathcal{V}-2 \hbar \omega \\
\end{array}
\right).
\end{equation}
where $\omega$ stands for a frequency of an external stimulus that induces a local transition in each subsystem, $\delta_{i}$ is the tunneling coupling strength in each coupled quantum dot, and the $\mathcal{V}$ represents the Coulomb interaction between the electrons. 
In such a scenario, the density matrix provides a simple approach to extend the investigation to finite temperatures when the system is in equilibrium with a thermal bath. In the canonical ensemble at temperature $T$, the thermal density matrix of the system can be written in the Gibbs form
\begin{eqnarray}
\varrho_T = \frac{1}{{Z}} \exp(-\beta H),
\label{eq:thermal_state}
\end{eqnarray}
\\
where $\beta=1/k_B T$ and $Z$ being the canonical partition function $Z= tr( e^{-\beta H})$. In this study, Boltzmann's constant $k_B$ and Planck's constant $\hbar$ are set to $1$. The thermal state, Eq.  \eqref{eq:thermal_state}, can be parameterize  in terms of the identity $\mathbb{I}$ and Pauli matrices $\{\sigma^{A(B)}_i\}$ as follows
\begin{eqnarray}
\label{eq:density1}
\varrho_T &=& \frac{1}{4} \big( \mathbb{I}_{AB}+ a_1 (T) \sigma^A _x \otimes \mathbb{I}_B + a_2 (T) \sigma^A _z \otimes \mathbb{I}_B \\ &+& \nonumber b_1 (T) \mathbb{I}_A \otimes \sigma^B _x + b_2 (T) \mathbb{I}_A \otimes \sigma^B _z + \sum_{i,j=1}^3 t_{ij} (T) \sigma^A _i \otimes \sigma^B _i\big)\nonumber,
\end{eqnarray}
where the non-vanishing thermal coefficients $a_{i} (T)$, $b_{i} (T)$ and $t_{ij} (T)$ can be written as
\begin{widetext}
\begin{eqnarray}
& a_1(T) = 2 \varrho_{13}+2 \varrho_{24}, \quad a_2(T) = \varrho_{11}+\varrho_{22}-\varrho_{33}-\varrho_{44}, 
\\
& b_1 (T) = 2 \varrho_{12}+2 \varrho_{34}, \quad  b_2(T) = \varrho_{11}-\varrho_{22}+\varrho_{33}-\varrho_{44},\\
& t_{11} (T) = 2 \varrho_{14}+2 \varrho_{23}, \quad t_{22} (T) = 2 \varrho_{23}-2 \varrho_{14}, \quad t_{33} (T) = \varrho_{11}-\varrho_{22}-\varrho_{33}+\varrho_{44}, \\
& t_{13} (T) = 2 \varrho_{13}-2 \varrho_{24}, \quad t_{31} (T) = 2 \varrho_{12}-2 \varrho_{34},
\end{eqnarray}
\end{widetext}
with $\varrho_{mn}$ being the matrix elements of the thermal state in the local basis ($\{\ket{i_{A},j_{B}}\}_{i,j=l,r}$).
The populations and the eigenvectors of the two coupled DQD system can be obtained numerically \footnote{The numerical results were obtained using functional programming (FP) paradigms using Julia programming language. FP produces numerically precise results with improved computational-time complexity due to the benefit of backward functions.} by diagonalizing the above density matrix. Figure \ref{popu} depicts the behavior of the system's populations as a function of temperature and Coulomb coupling in two scenarios: when $\omega < \mathcal{V}$  and when $\omega>\mathcal{V}$. In each scenario, a concrete comparison with the absence of the stimulus effect is presented. 
\begin{widetext}
\begin{minipage}{\linewidth}
\begin{figure}[H]
\centering
\subfigure[]{\label{Figp_a}\includegraphics[scale=0.59]{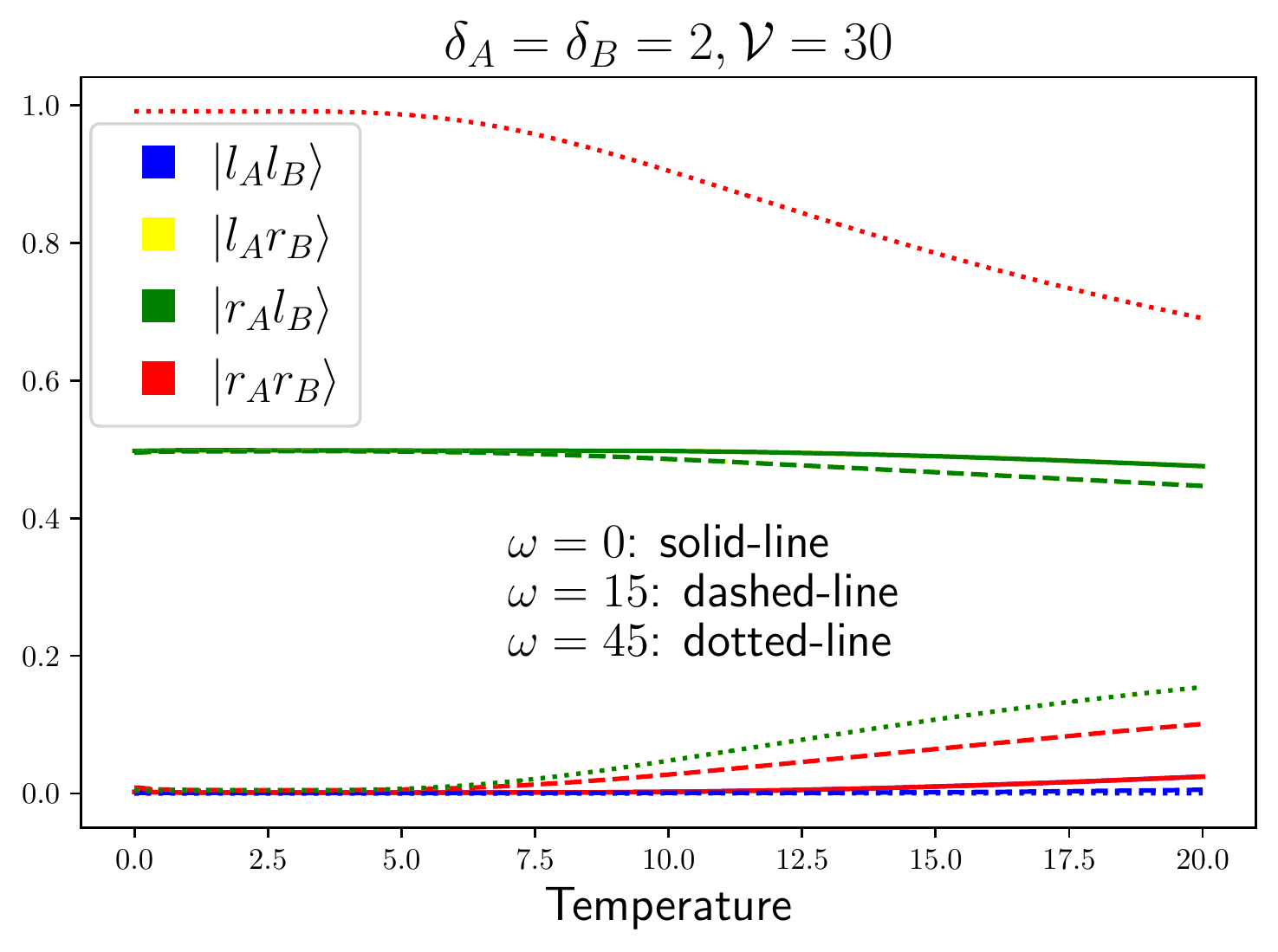}}
\subfigure[]{\label{Figp_b}\includegraphics[scale=0.59]{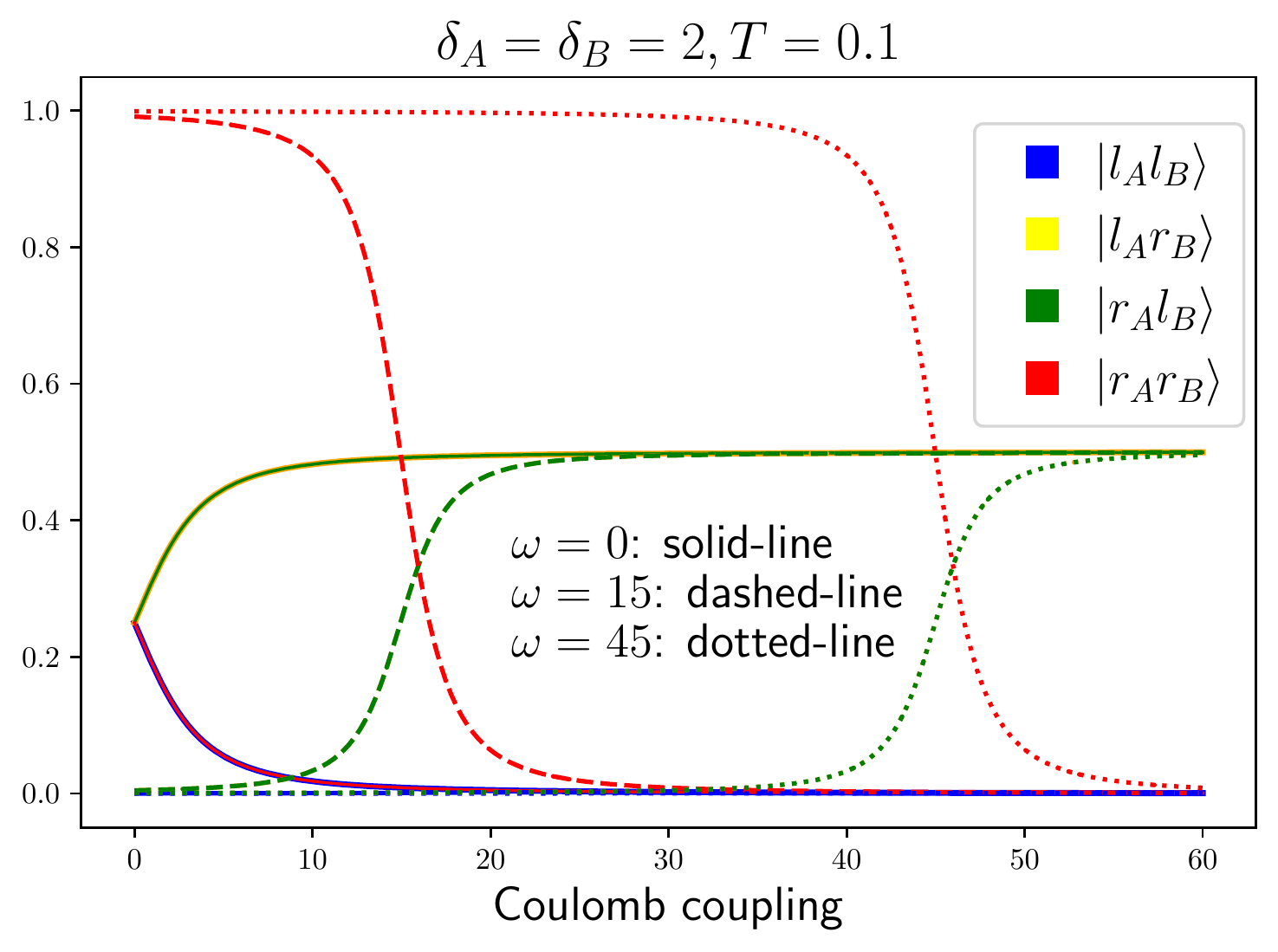}}
\caption{Comparison of the system's populations with and without the external stimulus effect  ($\omega = 0, 15$, and $45$).  \ref{Figp_a}: versus temperature. \ref{Figp_b}: versus Coulomb coupling. 
}
\label{popu}
\end{figure}
\end{minipage}
\end{widetext}
As can be seen, in the absence of the external stimulus, there is no energy level crossing, thus, no population inversion. In this scenario, the effect of the Coulomb coupling leads the system to the configuration described by the maximally entangled state $\left( \ket{r_A l_B}+\ket{l_A r_B}\right)/\sqrt{2}$.  However, when the external stimulus is turned on, it splits the energy levels in such a way that both electrons are found in the state $\ket{r_A r_B}$ for $\omega > \mathcal{V}$. Thus, the introduction of the external stimulus creates a dispute between its energetic contribution and the Coulomb coupling, which leads to a quantum level crossing between the separable state $\ket{r_A r_B}$ and the maximally entangled one  $\left( \ket{r_A l_B}+\ket{l_A r_B}\right)/\sqrt{2}$, characterized by the threshold point $\mathcal{V}_c = \omega$. As can be seen, increasing the strength of the Coulomb coupling will suppress the external stimulus, leading the system to the state $\left( \ket{r_A l_B}+\ket{l_A r_B}\right)/\sqrt{2}$. This quantum level crossing between a separable and an entangled state plays an important role in the behavior of the quantum features of the system, as shown in section \ref{sec4}.

\section{Quantum information quantifiers \label{sec2}}

In the following, we introduce the quantum information quantifiers used to characterize the quantum coherence and entanglement in the thermal state of two coupled DQDs in the presence of an external stimulus. 

\subsection{\texorpdfstring{$l_1$}--norm coherence} 

Quantum coherence is a basis-dependent feature of a quantum state which emerges from the superposition of the quantum states of the system, represented by the off-diagonal elements of the density matrix $\varrho$, Eq. \eqref{eq:density1}, in a certain basis \cite{MK,R40}. Baumgratz $et~ al.$ \cite{R40} has proved that  $l_1$-norm is a reliable coherence measurement and is effortlessly computed for a given density matrix $\varrho$ as
\begin{equation}
\mathcal{C}_{l_{1}}(\varrho)=\sum_{i\neq j}|\varrho_{ij}|\label{eq:norm-l1}.
\end{equation}
It is worth noting that quantum coherence, distinct from other quantum correlation quantifiers, gives additional information about a quantum system's essential nature. Furthermore, while quantum correlations needs the definition of at least two parties, quantum coherence may be established for just a sole qubit system.  On the other hand, for any given multipartite state $\varrho$, the correlated contribution in the quantum coherence can computed by taking away the local coherence of subsystems $\varrho_A$ and $\varrho_B$ from the total coherence \cite{tan2018,R51,kraft2018}. The reduced density matrix of subsystem $A$ is given by
\begin{eqnarray*}
\varrho_T ^A &=& (\varrho_{11} +\varrho_{33}) \ket{l_A} \bra{l_A} + (\varrho_{22} 
+ \varrho_{44}) \ket{r_A} \bra{r_A}\\ &+& (\varrho_{13} +\varrho_{24}) (\ket{l_A} \bra{r_A} + \ket{r_A} \bra{l_A}).
\end{eqnarray*}
Similarly, for subsystem $B$, we obtain
\begin{eqnarray*}
\displaystyle
\varrho_T ^B &=& (\varrho_{11} +\varrho_{33}) \ket{l_B} \bra{l_B} + (\varrho_{22} 
+ \varrho_{44}) \ket{r_B} \bra{r_B}\\ &+& (\varrho_{12} +\varrho_{34}) (\ket{l_B} \bra{r_B} + \ket{r_B} \bra{l_B}) .
\end{eqnarray*}
Thus, the definition of correlated coherence based on the $l_1$-norm of coherence is defined as
\begin{equation}
\mathcal{C}_{cc} (\varrho) := \mathcal{C}_{l_1} (\varrho)-\mathcal{C}_{l_1} (\varrho_A)-\mathcal{C}_{l_1} (\varrho_B),
\end{equation}
where the density matrices of local subsystems are $\varrho_A = tr_B \varrho$ and  $\varrho _B = tr_A \varrho$. 
In the case of two coupled DQDs, the inclusion of the external stimulus contribution renders the analytical formulation for the eigenvalues of the Hamiltonian and the density matrix entries too complicated to be fully stated within this text. For the sake of simplicity, using the implicit form of the density matrix in Eq. \eqref{eq:density1}, one can straightforwardly compute the total ($\mathcal{C}_{l_{1}}^{T}$), local ($\mathcal{C}_{l_{1}}^{L}$) and correlated ($\mathcal{C}_{l_{1}}^{C}$) ${l_{1}}$-norm based quantum coherence:
\begin{eqnarray}
\mathcal{C}_{l_{1}}^{T}(\varrho_T) &=& 2 \Big(\left|\varrho_{12}\right| + \left|\varrho_{13}\right| + \left|\varrho_{14}\right| \nonumber\\
&+& \left|\varrho_{23}\right| + \left|\varrho_{24}\right| + \left|\varrho_{34}\right|\Big), 
\\
\mathcal{C}_{l_{1}}^{L}(\varrho_T) &=& \left|a_1 (T)\right| + \left|b_1 (T)\right|\nonumber\\
&=& 2 \Big(\left|\varrho_{13} + \varrho_{24}\right| + \left|\varrho_{12} + \varrho_{34}\right|\Big), 
\\
\mathcal{C}_{l_{1}}^{C} (\varrho_T) &=& 2 \Big(\left|\varrho_{12}\right| + \left|\varrho_{13}\right| + \left|\varrho_{14}\right| + \left|\varrho_{23}\right| + \left|\varrho_{24}\right| \nonumber\\
&+& \left|\varrho_{34}\right|  - ( \left|\varrho_{13} + \varrho_{24}\right| + \left|\varrho_{12} + \varrho_{34}\right|)\Big). 
\end{eqnarray}
In Sec. \ref{sec4}, we give and discuss the main findings concerning quantum correlations and quantum coherence simulation in terms of temperature and the different parameters appearing in the given Hamiltonian of the system.

\subsection{Thermal entanglement}

In order to make a comparison between the amount of entanglement and the coherence in the system, we adopt the measurement of concurrence, which is typically used to quantify the entanglement in bipartite systems \cite{hill1997entanglement,wootters2001entanglement}. The thermal concurrence explores the similarity between the  state of the system under consideration in thermal equilibrium, Eq. \eqref{eq:density1}, and its  bit-flipped density matrix, defined as:
\begin{equation}
R = \varrho_T (\sigma^y \otimes \sigma^y)  \varrho_T ^{*} (\sigma^y \otimes \sigma^y),
\end{equation}
with $\sigma^y$ being a Pauli matrix. Thus, Concurrence is then defined as
\begin{equation}
\mathcal{C} (\varrho_T) := \rm{max} \{0, \sqrt{\lambda_1}-\sqrt{\lambda_2}-\sqrt{\lambda_3}-\sqrt{\lambda_4} \},
\end{equation}
where $\lambda_i$'s are sorted in decreasing order, $\lambda_1 > \lambda_2 \geq \lambda_3 \geq \lambda_4$, to ensure concurrence positivity. In the following Sec. \ref{sec4}, concurrence is applied to to analyze the effect of an induced transition on the quantum entanglement of a two coupled DQDs system.
\\
\section{Results and discussion} \label{sec4}
In order to understand the effect of an induced transition on the quantum coherence and entanglement of our system, we need to analyze how the quantum information quantifiers change as a function of each parameter of the Hamiltonian. Thus, we dedicate this section to presenting and discussing the behavior of the quantum features presented in Section \ref{sec2}, associated with the change of the transition frequency, Coulomb potential, and tunneling strengths. Then, in light of the findings, we present a comprehensive analysis of outcomes and assess the impact of a transition induced by an external stimulus on the proposed model.

As our first analysis, we keep the Coulomb coupling and the tunneling coupling strength fixed as we allow the temperature to increase for selected frequencies. Fig. \ref{Fig1} shows the quantum information quantifiers as a function of the temperature for different values of transition frequency with fixed Coulomb coupling and tunneling strengths.

The effect of the external stimulus is responsible for the transition between the right and left dots, i.e., the higher its frequency higher is the probability of transition from one dot to another. On the other hand, this effect also depends on the values of the other parameters, i.e., there is a competition on the effect of the Coulomb, tunneling coupling strength, and the transition frequency. This competition leads to a change in the occupation of the energy levels, as presented in Fig. \ref{Figp_b}.  As a consequence, increasing the transition frequency will change the degree of correlation in the system, as depicted in Fig. \ref{Fig1}.

As it can be seen, increasing the frequency above the quantum level crossing $\omega > \mathcal{V}$ (\ref{Figp_b}) will populate many incoherent separable states, decreasing the degree of coherence and entanglement of the system.  
On the other hand, for frequencies lower than the Coulomb coupling ($\omega < \mathcal{V}$), at low temperatures, the increasing in the transition frequency may lead the system to some excited states with a larger degree of coherence and entanglement, yielding a enhancement on the quantum features. 

As expected, the effect will become indistinguishable as we increase the temperature. Thus, for higher values of temperature, the quantum features of the system tend to fade, and the correlation between both electrons is destroyed. In particular, for a given critical temperature $T_c$, where there is a sudden death to the concurrence (entanglement). However, from the correlated coherence, one can verify that this system remains correlated despite the absence of entanglement. This result strengths the fact that correlated coherence in DQD system is more robust than the thermal entanglement \cite{R51}.  On the other hand, as correlated coherence goes to zero for higher temperatures, the local coherence is still greater and becomes the main contribution to the quantum coherence of the system. Thus, in high-temperature regimes, one can still observe some coherence from the local contribution, even in the absence of correlations between the electrons.
\begin{widetext}
\begin{minipage}{\linewidth}
\begin{figure}[H]
\centering
\subfigure[]{\label{Fig1_a}\includegraphics[scale=0.55]{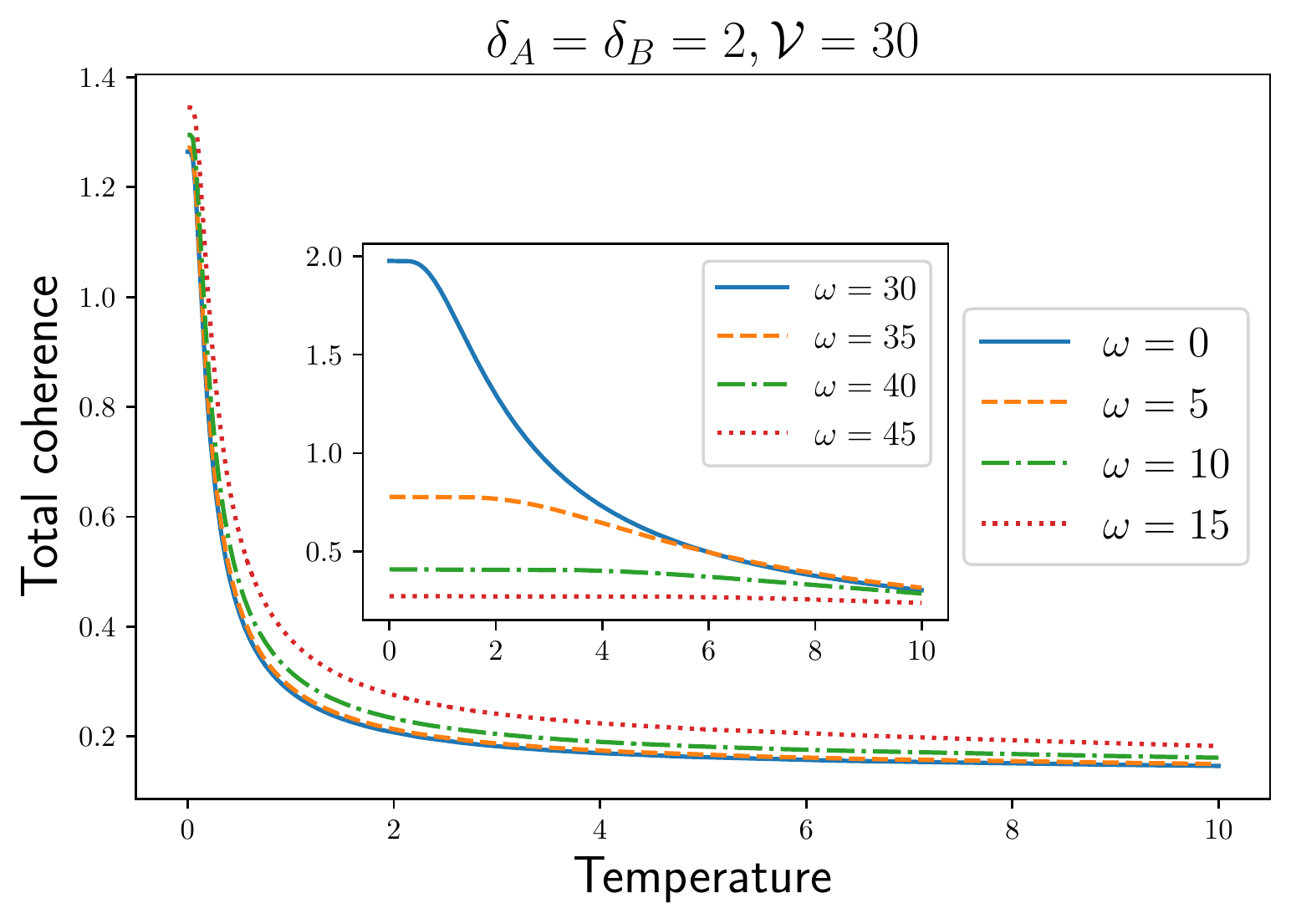}}
\subfigure[]{\label{Fig1_b}\includegraphics[scale=0.55]{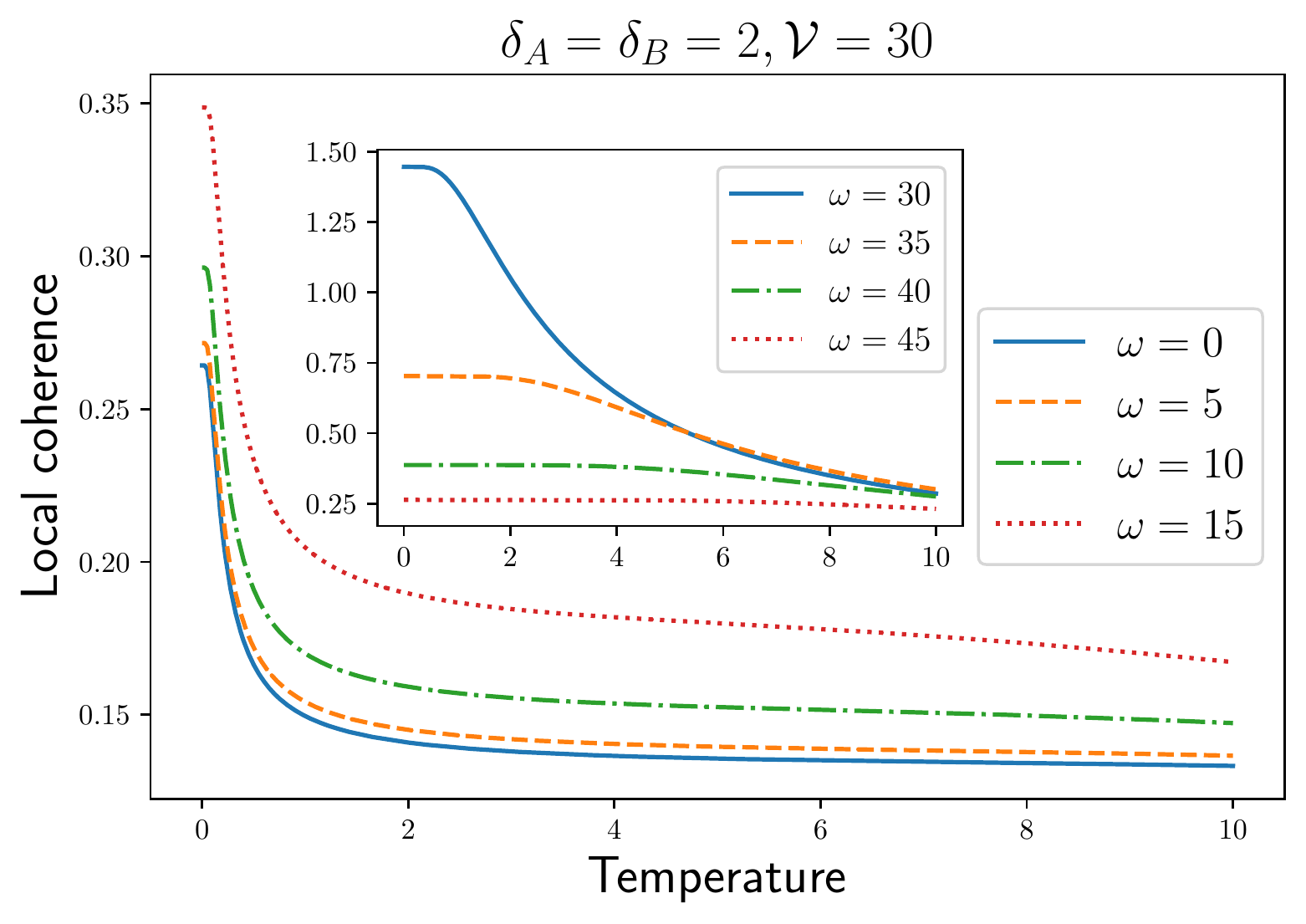}}
\subfigure[]{\label{Fig1_c}\includegraphics[scale=0.55]{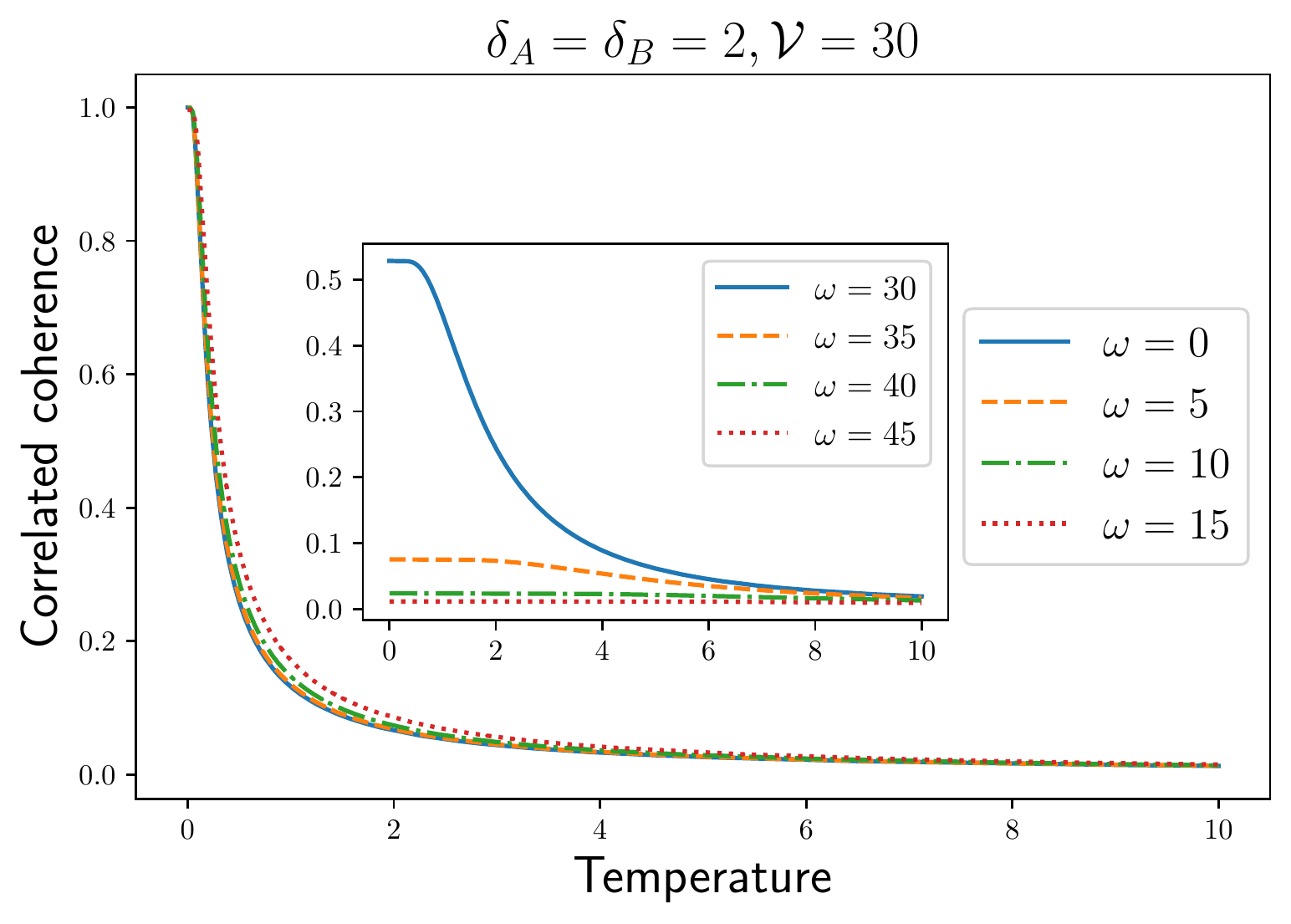}}
\subfigure[]{\label{Fig1_d}\includegraphics[scale=0.55]{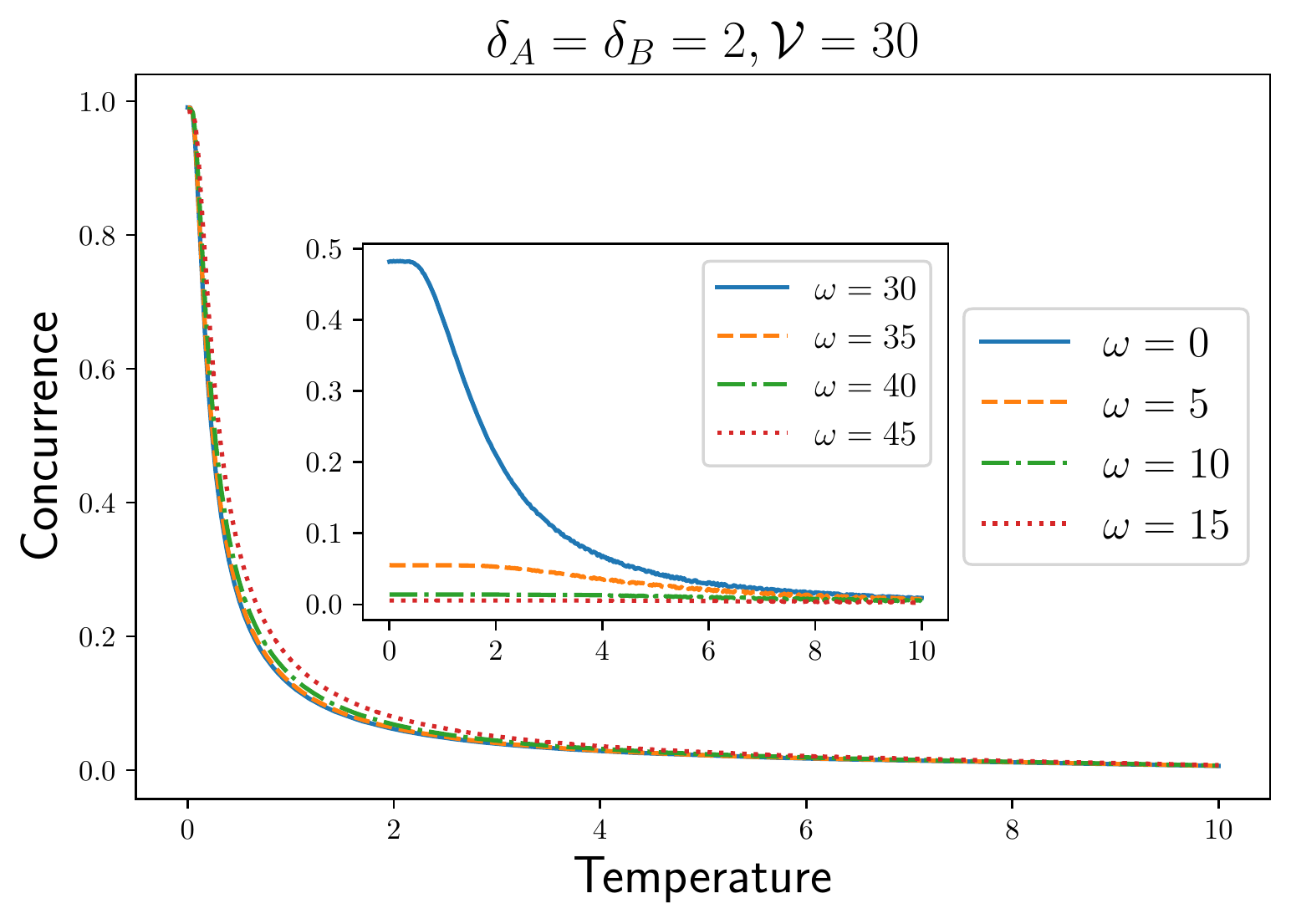}}              
\caption{Total coherence \ref{Fig1_a}, local coherence \ref{Fig1_b}, correlated coherence \ref{Fig1_c} and concurrence \ref{Fig1_d} as a function of the temperature for different values of transition frequency $\omega < \mathcal{V}$ and $\omega > \mathcal{V}$ (inset), with $\delta_{A}=\delta_{B}=2$ and $\mathcal{V} = 30$.}
\label{Fig1}
\end{figure}
\end{minipage}
\end{widetext}

In the following, we keep the system at the low temperature of $T = 0.1$ while the Coulomb coupling is allowed to change. At this low value of temperature, we find our system with a high degree of coherence and concurrence, according to Fig. \ref{Fig1}. On the other hand, Fig. \ref{Fig2} shows the dependence of the quantum information quantifiers as a function of the  Coulomb coupling. As can be seen from total and local coherence, for non-zero transition frequency, a first increase in the Coulomb coupling strength enhances quantum resources within the system. However, after it exceeds a threshold value $\mathcal{V}_c = \omega$, these quantifiers decrease as we increase the Coulomb coupling. This behavior is explained by the same mechanism presented in Fig. \ref{Figp_b}. For $\mathcal{V}_c > \omega$ occurs a population inversion due to the energy levels crossing due to the competition between the Coulomb coupling strength and the transition frequency, as observed in Fig.\ref{popu} and \ref{Fig1}.

\begin{widetext}
\begin{minipage}{\linewidth}
\begin{figure}[H]
\centering
\subfigure[]{\label{Fig2_a}\includegraphics[scale=0.55]{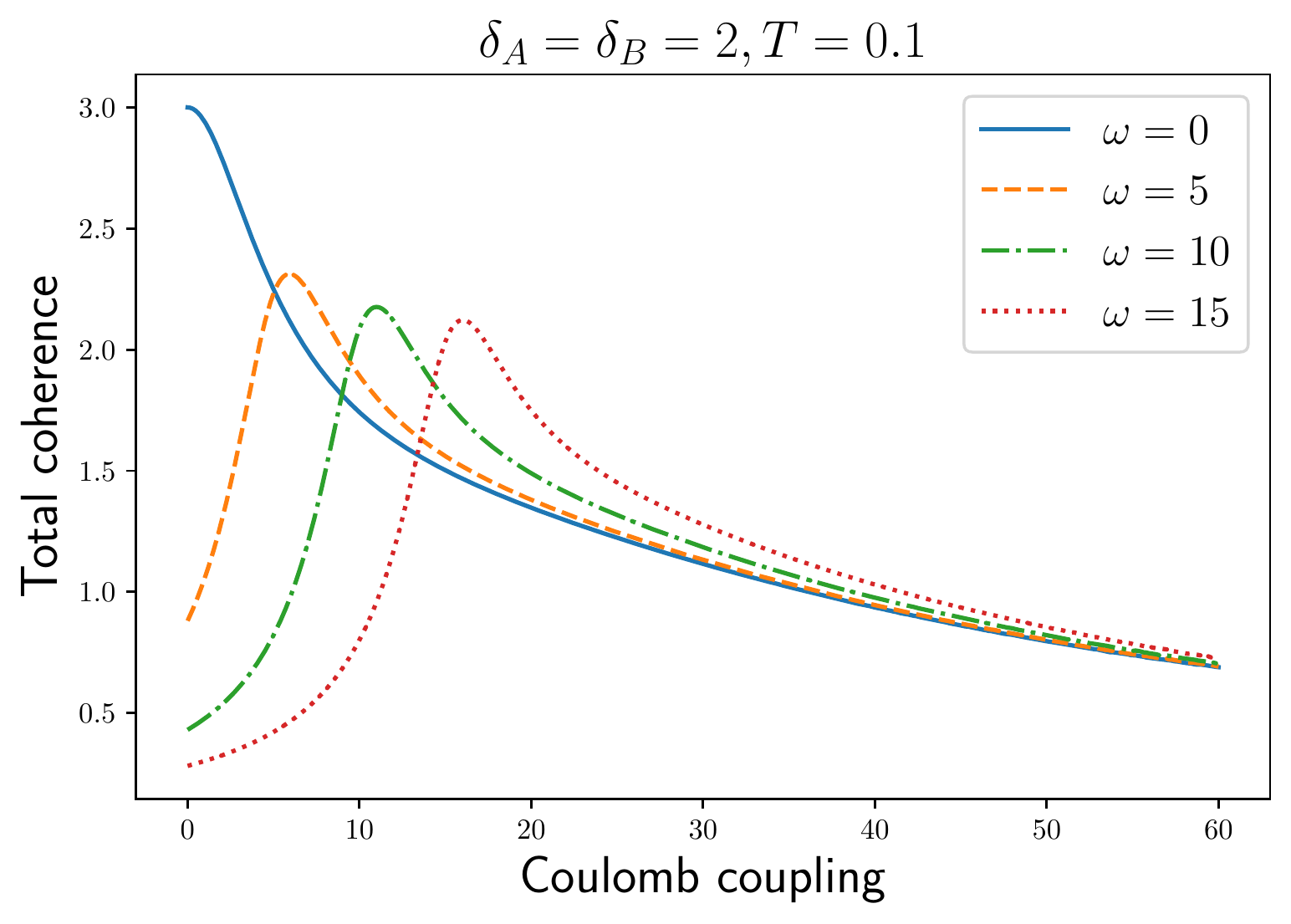}}
\subfigure[]{\label{Fig2_b}\includegraphics[scale=0.55]{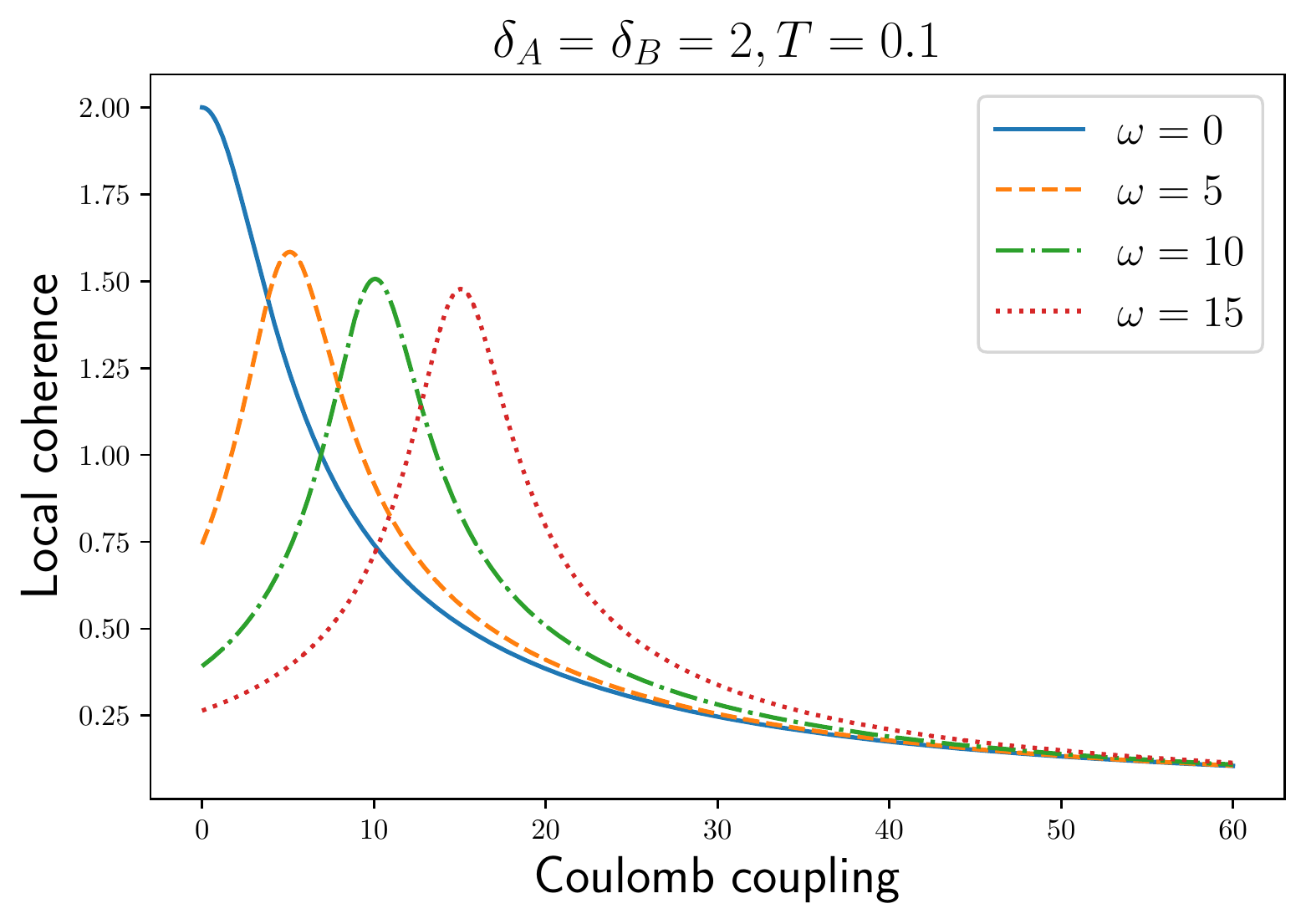}}
\subfigure[]{\label{Fig2_c}\includegraphics[scale=0.55]{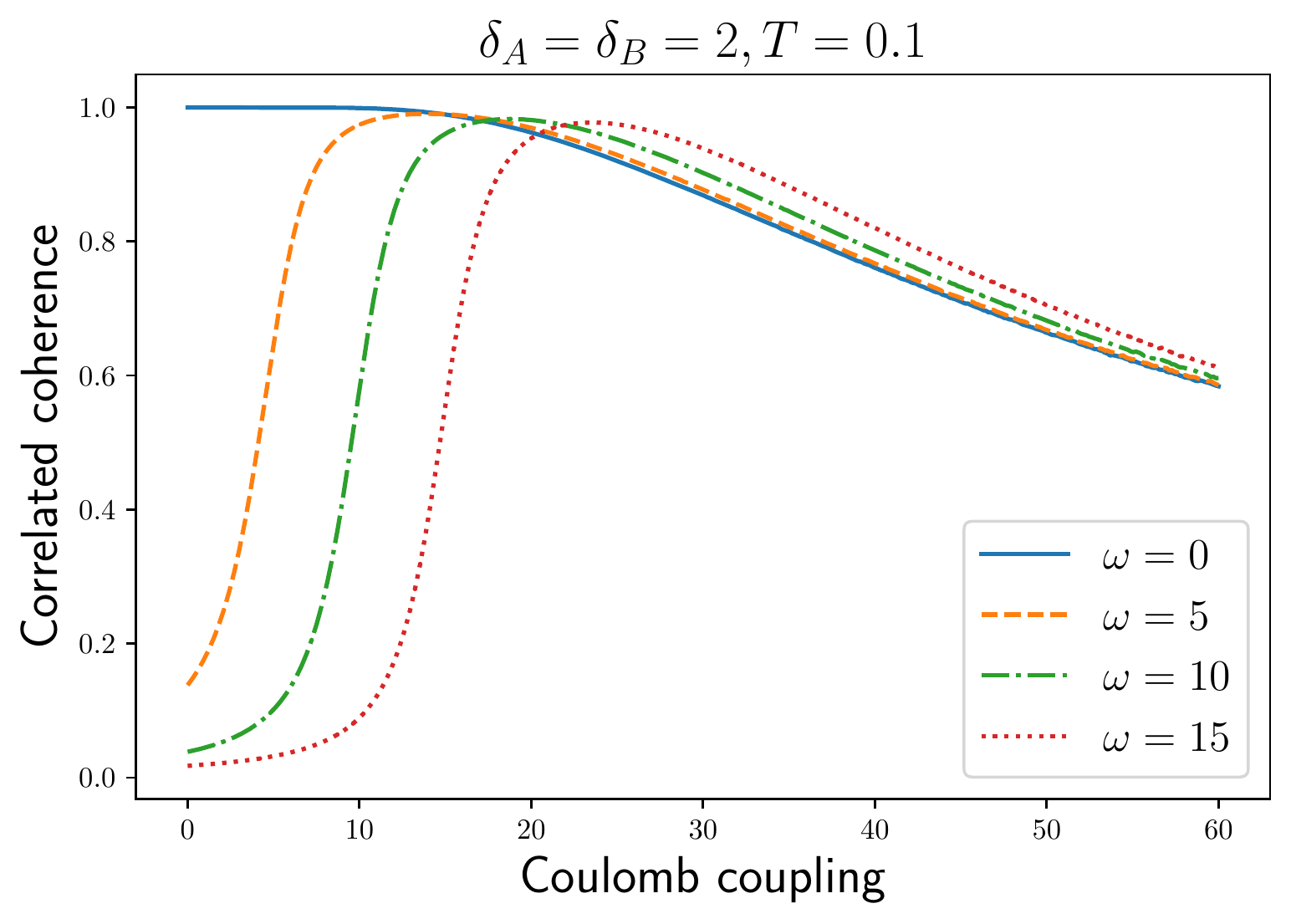}}
\subfigure[]{\label{Fig2_d}\includegraphics[scale=0.55]{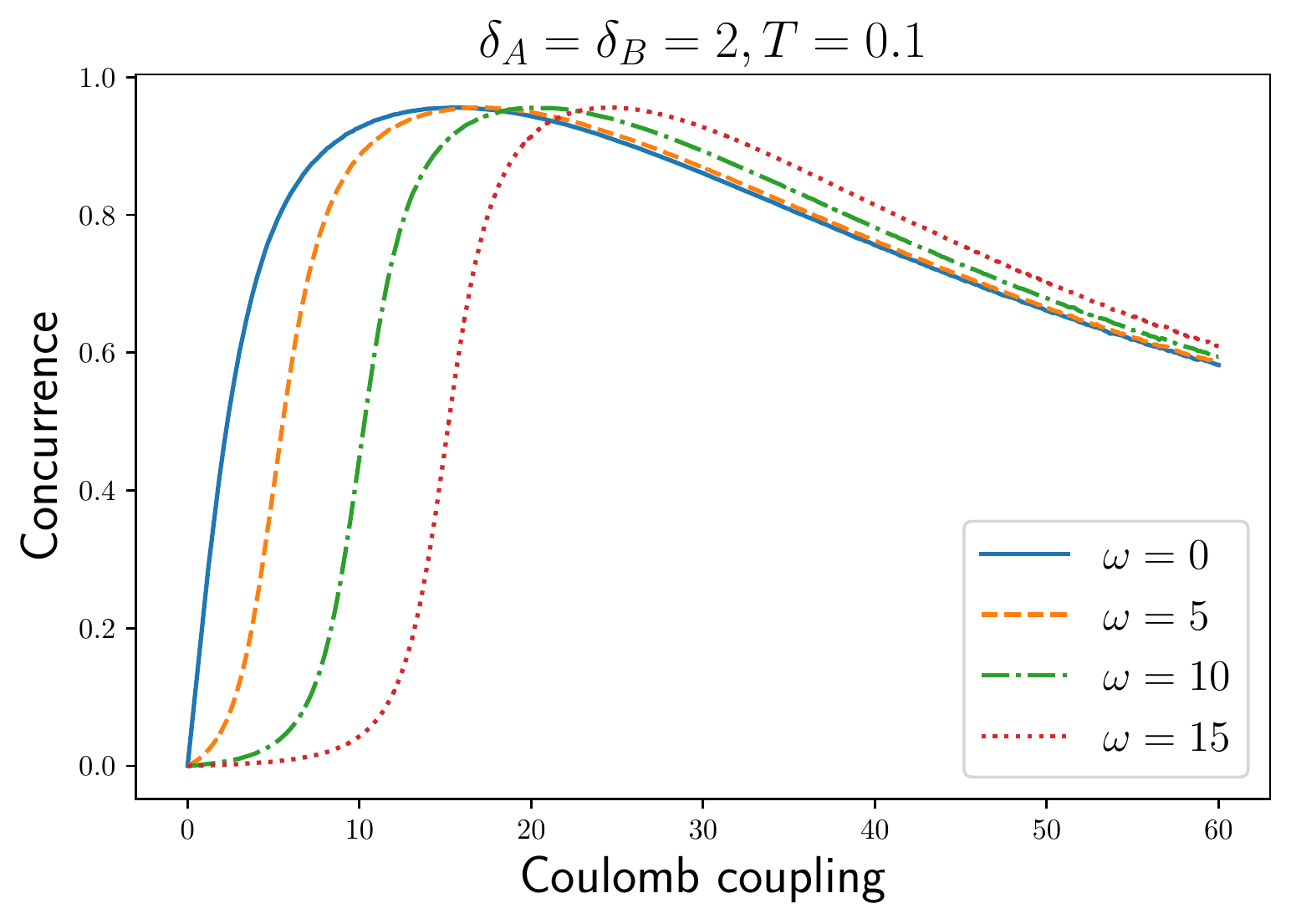}}              
\caption{Total coherence \ref{Fig2_a}, local coherence \ref{Fig2_b}, correlated coherence \ref{Fig2_c} and concurrence \ref{Fig2_d} as a function of the Coulomb coupling $\mathcal{V}$ for different values of  transition frequency, with $\delta_{A}=\delta_{B}=2$ and $T = 0.1$.}
\label{Fig2}
\end{figure}
\end{minipage}
\end{widetext}

It is worth noting that the concurrence is zero for $\mathcal{V} = 0$. This effect is expected since the Coulomb coupling is responsible for the interaction between both electrons. On the other hand, the correlated coherence goes to a finite non-zero value, reaching its maximum for $\omega = 0$. Thus, the correlation contribution to the coherence is not all due to the presence of entanglement. 

Furthermore, as we increase the Coulomb coupling, we change the predominant contribution for the total coherence from the local to the correlated coherence. This effect relies on the fact that the Coulomb coupling increases the interaction between both electrons in the different dots while decreasing the superposition generated by the transition frequencies.

In Fig. \ref{Fig3}, we plot the quantum information quantifiers as a function of the tunneling strengths $\delta_A = \delta_B$ for different values of transition frequency.  As can be seen,  different from Fig. \ref{Fig1}, we find that the behavior of the correlated coherence and the concurrence does not match. We observed a sudden increase in the entanglement for weak values of the tunneling parameter. After reaching a maximum value, the entanglement gradually decreases to zero as we increase the tunneling strengths. 
On the other hand, as we increase the tunneling strength, the coherence emerging from the superposition reaches its maximum value, which indicates the presence of pure states in this regime. In addition, it can be seen that higher transition frequencies rapidly drive the state of the system from being a mixed entangled state to a pure state with a lower degree of entanglement.  

\begin{widetext}
\begin{minipage}{\linewidth}
\begin{figure}[H]
\centering
\subfigure[]{\label{Fig3_a}\includegraphics[scale=0.55]{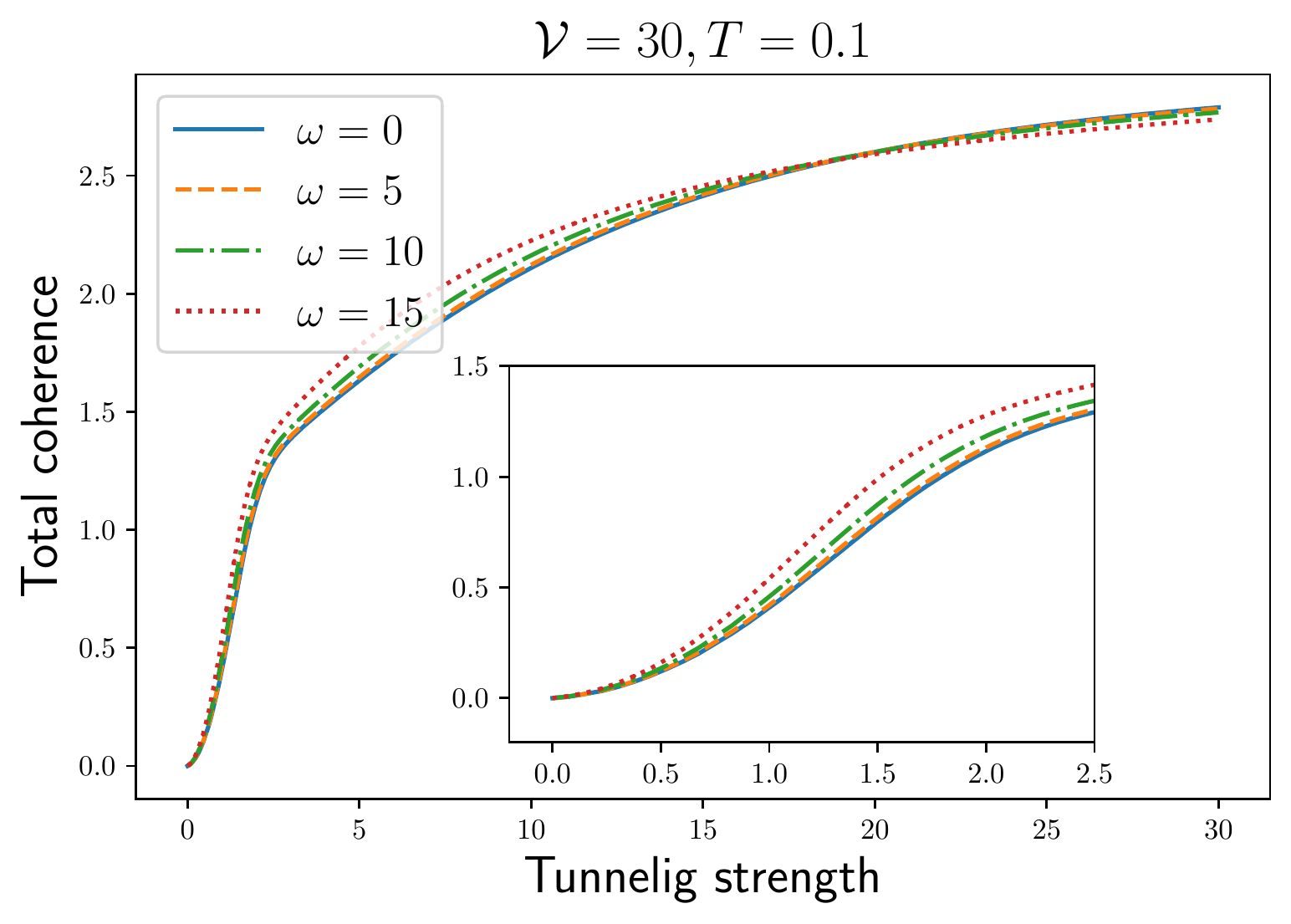}}
\subfigure[]{\label{Fig3_b}\includegraphics[scale=0.55]{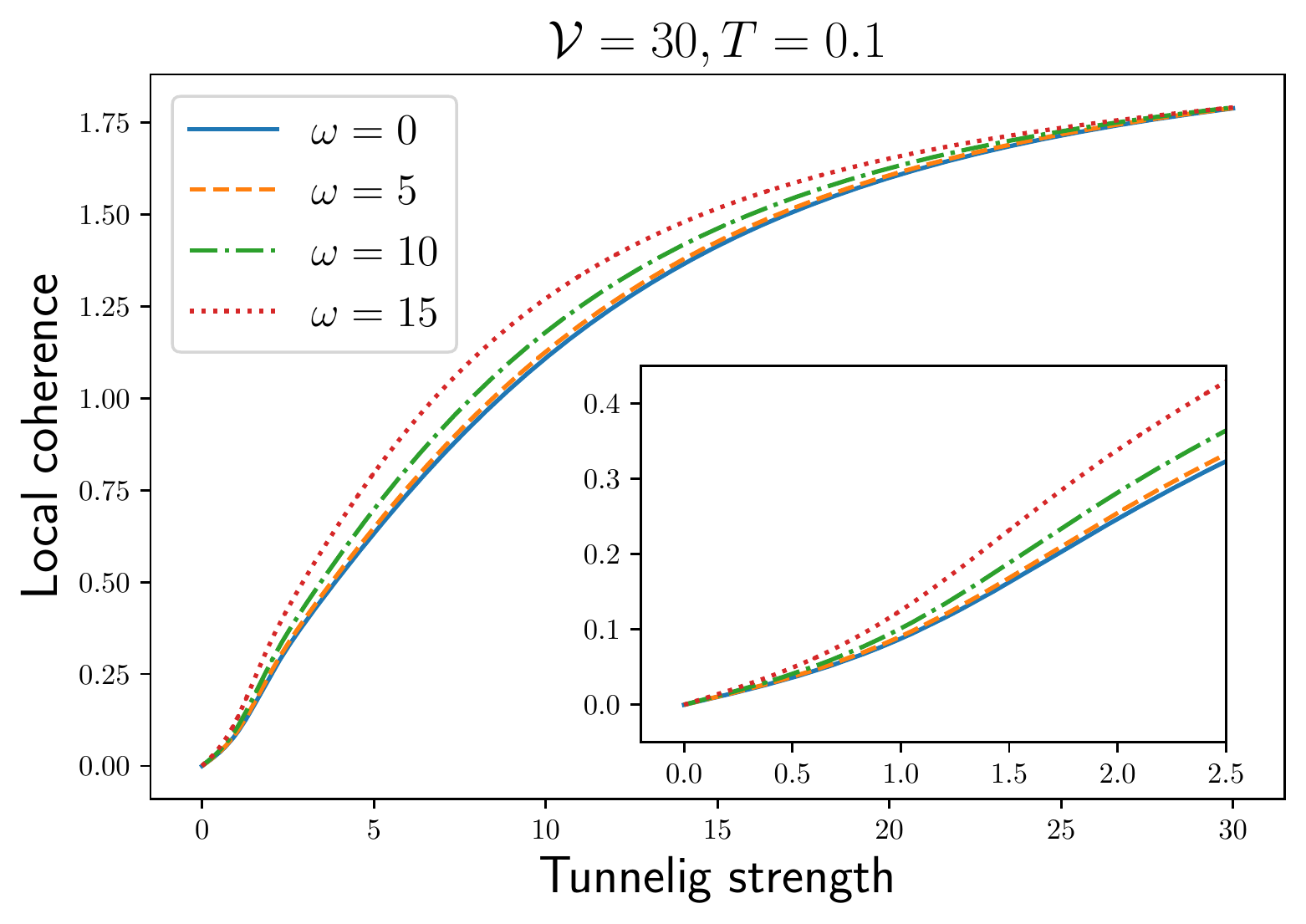}}
\subfigure[]{\label{Fig3_c}\includegraphics[scale=0.55]{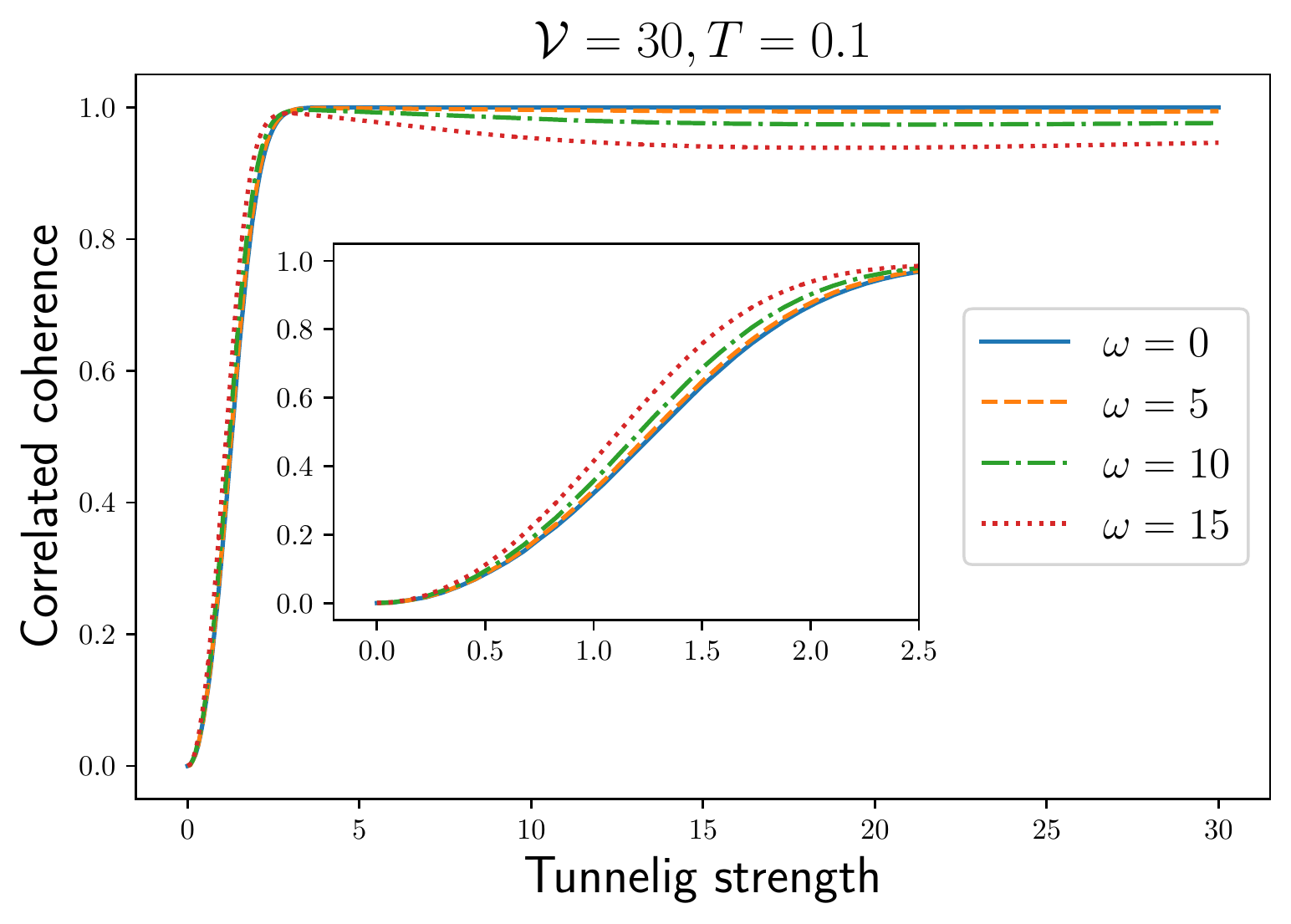}}
\subfigure[]{\label{Fig3_d}\includegraphics[scale=0.55]{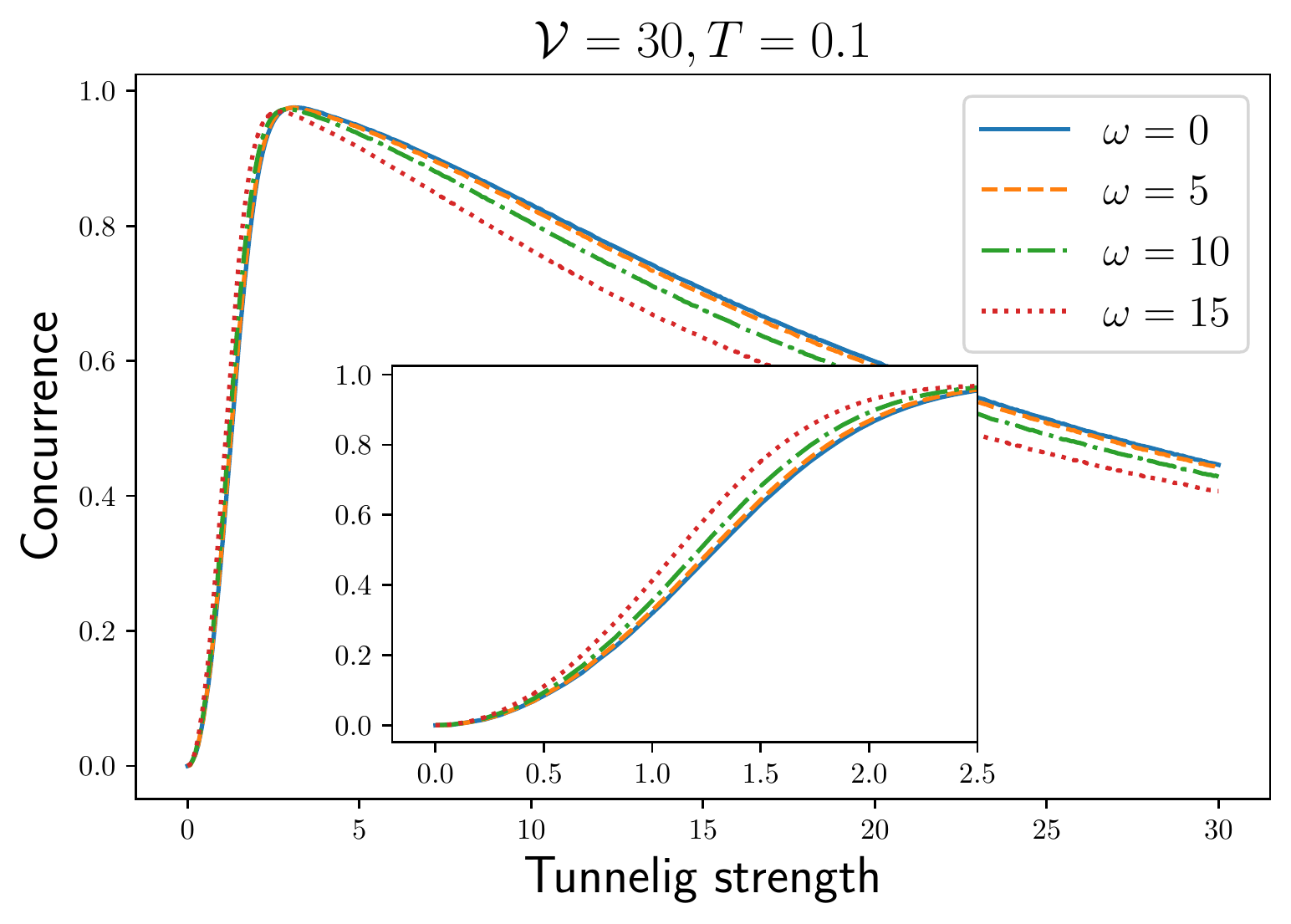}}              
\caption{Total coherence \ref{Fig3_a}, local coherence \ref{Fig3_b}, correlated coherence \ref{Fig3_c} and concurrence \ref{Fig3_d} as a function of the tunneling strengths $\delta_A = \delta_B$ for different values of  transition frequency, with $\mathcal{V} = 30$ and $T = 0.1$. Inset shows the enhancement of the quantum coherence and entanglement due to the increase of the transition frequency for $0< \omega < \mathcal{V}$}
\label{Fig3}
\end{figure}
\end{minipage}
\end{widetext}

\section{Conclusions}

In this work, we analyzed the effect of an induced transition, yielded by an external stimulus, on the quantum features of a two-coupled double quantum dots system by quantifying the thermal quantum coherence and entanglement. The quantifiers were defined based on the formulated resource theory of coherence and the measurement of concurrence. The physical model is numerically solved in order to obtain the thermal density operator, which allows the evaluation of the quantum information quantifiers. The introduction of an external stimulus induces a quantum-level crossing on the populations of the system due to the competition between the energetic contribution of the stimulus and the Coulomb coupling of the system. The results have shown that this competition plays a pivotal role in the quantum properties of the system, leading to the enhancement of the entanglement and the coherence for $0< \omega < \mathcal{V}$. In this regard, the transition frequency can be tuned in order to handle the quantum properties of the two coupled DQDs systems. These results provide insight into the quantum properties of DQDs and contribute to a better understanding of these properties leading to promising applications in semiconductor nanostructures such as double quantum dots.

\section*{Acknowledgement}
ZD would like to thank the Abdus Salam International Centre for Theoretical Physics (ICTP) for their hospitality and access to their research facilities during his stay, which assisted in part of this work. Maron F. Anka thanks Coordenação de Aperfeiçoamento de Pessoal de Nível Superior - Brasil (CAPES) - Finance Code 001, and the Carlos Chagas Filho Foundation for Research Support of Rio de Janeiro (FAPERJ) for financial support. MR acknowledges support from National Council for Scientific and Technological Development (CNPq) - Grant No. 317324/2021-7.

\bibliographystyle{ieeetr}

\bibliography{article.bib}

\end{document}